\documentclass[aps,twocolumn,showpacs,aps]{revtex4}
\usepackage{graphicx}
\usepackage{dcolumn}
\usepackage{amsmath}
\usepackage{amssymb}

\begin{document}

\title{Reduction in the electron density-of-states in superconducting MgB$_2$ 
disordered by $n^1$-irradiation: the $^{11}$B and $^{25}$Mg NMR estimates}

\author{~A.~P.~Gerashenko, K~.N.~Mikhalev, ~S.~V.~Verkhovskii, ~A.~E.~Karkin,
 ~B.~N.~Goshchitskii}
\affiliation{Institute of Metal Physics, UB RAS, 
             620219 Ekaterinburg GSP-170,Russia}

\smallskip 
\begin{abstract}

NMR line shift and nuclear spin-lattice relaxation rate $T_1^{-1}$ of
$^{11}$B and $^{25}$Mg were measured in superconducting MgB$_2$
($T_c^{ons}$=38K) structurally disordered by nuclear reactor neutrons up to
the fluence of thermal neutrons $\Phi=1 \cdot 10^{19}$~cm$^{-2}$.
The temperature of superconducting transition was shifted down to
$T_{c,irrad}^{ons}$=7K under irradiation. The change due irradiation 
in the partial electron density of states (DOS) at the Fermi energy of boron 
and magnesium  were traced by taking into account that $T_1^{-1}$ of $^{11}$B and 
$^{25}$Mg are determined by hyperfine magnetic interactions with carriers. It was   
revealed that electronic states near $E_F$ of Mg are influenced negligibly by 
irradiation whereas partial DOS of the B $2p_{x,y}$ states reduces greatly in 
irradiated MgB$_2$. Acording to the Mc Millan formula, NMR data show 
that critical temperature decreases in irradiated MgB$_2$ mainly due to reduction in 
the partial DOS of $p$ states of boron.
\end{abstract}


\maketitle
\smallskip
\section{Introduction}

Discovery of superconductivity in magnesium diboride MgB$_2$ showing  
high temperature of superconducting transition ($T_c \sim 40$K)
\cite{Akimitsu} stimulated extensive studies of its electron properties.
Layered crystal structure of MgB$_2$ (space group symmetry $D_{6h}^1$)
is well known and relates to the hexagonal AlB$_2$-type ~\cite{Goldschmidt}.
Boron forms a primitive honeycomb lattice and pronounced B-B covalent
bonding creates graphite-like sheets of boron separated by hexagonal layer
of Mg and interatomic bonding is much more metallic in origin. Recent 
first principles calculations of electronic band structure show
that the bands near the Fermi energy are formed mainly from $2p_{x,y}$
bonding orbitals of boron ~\cite{Kortus,Medvedeva,An}.
As predicted the dispersion of these bands is extremely small near
the $\Gamma$ point of the Brillouin zone. These bands form two 
small cylindrical Fermi surfaces around $\Gamma$-A line. Due to their
2D character they contribute near the Fermi energy more than one third
of the total DOS - $N(2p_{x,y})$=0.072(eV$\cdot$spin$\cdot$at.B)$^{-1}$
~\cite{Belashenko}. Sizeable electron-phonon coupling
$\lambda\sim 0.75-0.87$ is predicted for electrons in these 
$2p_{x,y}$ bands and strong boron isotope effect reported in ~\cite{Bud'ko}
is in favor that MgB$_2$ being phonon mediated superconductor.
The parallels regarding to band structure of MgB$_2$ and intercalated
graphite were enlightened in ~\cite{An,Belashenko_2}. An interplay 
between 2D covalent in-plane ($\sigma$) and 3D metallic-type ($\pi$) conducting bands accompanied by 
corresponding ($\sigma\leftrightarrow \pi$) charge transfer was considered as
important factor determining electronic properties and superconductivity
in MgB$_2$. 

The influence of structural disorder on the electron states in conduction band of 
MgB$_2$ is question of great interest both in physics and technology. As known 
high-pressure technology in sintering of the compacted MgB$_2$-materials results in 
stressed crystal structure accompanied by decrease of $T_c$ and broad 
superconducting transition in comparison with as grown at ambient pressure crystals. 
Radiation disordering by neutrons is probably the purest method to study the 
influence of the induced structural disorder on the physical properties regarding to the 
motion of carriers in conduction band.  As shown in  ~\cite{Karkin} superconducting 
temperature $T_c$ drops below 10K whereas the initial crystal structure  is  preserved 
under $n^1$ -irradiation up to  the fluence of thermal neutrons $\Phi=1\cdot 10^{19}$cm$^{-2}$. 
The moderate irradiation leads to anisotropic expansion of the crystal lattice with 
increase of the $c/a$ ratio resulting in increase of the interlayer distance. The Rietveld 
analysis refinement of X-ray diffraction patterns yields some decrease in the 
occupation number at Mg-sites. 

NMR line shift and nuclear spin-lattice relaxation rate (NSLRR) $T_1^{-1}$ 
studies give an unique opportunity to probe electronic states near $E_F$ through static 
and fluctuating parts of hyperfine magnetic fields created by carriers at the nuclei. 
Near a given NMR probe the wave function of carriers might be considered as the 
atomic-like. This permit to analyze the orbital content of the carrier states and to 
estimate contributions of different atoms to the total DOS near $E_F$. Similar detailed 
consideration of the experimental data regarding $T_1^{-1}$ of $^{11}$B in
structurally ordered MgB$_2$ \cite{Kotegawa,Gerashenko,Jung} was presented in
\cite{Belashenko,Pavarini} on the basis of band 
calculations and it was concluded that the orbital hyperfine interaction  of $2p$-holes  
with nuclear magnetic moments should  dominate in NSLRR of $^{11}$B. This 
contribution to $^{11}T_1^{-1}$ is proportional to ${N(2p_{x,y})}^2$ and depends on the 
orbital content of $2p_i$ electrons, having appropriate DOS 
$N_i=f_i*N(E_F)$ near the Fermi energy \cite{Obata,Asada}.
As shown in \cite{Pavarini} the Fermi-contact interaction should
be responsible for the Knight shift and NSLRR of $^{25}$Mg in MgB$_2$. 

We have measured the NMR line shift and nuclear spin-lattice relaxation rate 
of $^{11}$B and $^{25}$Mg in the ordered unirradiated
($T_c^{onset}$=38K) and disordered by 
$n^1$-irrradiation ($T_c^{onset}$=7K) MgB$_2$ to study the change
in DOS of the conduction bands. 

\section{ Experimental results and discussion}

Irradiation of polycrystalline sample was performed in nuclear reactor IVV-2M 
at $T\sim 350$~K up to the fluence of thermal neutrons
$\Phi=1\cdot 10^{19}$~cm$^{-2}$, and of fast neutrons,
to $5\cdot 10^{18}$~cm$^{-2}$ (total dose - of more than 10 dpa).
Superconducting transitions measured by AC susceptibility for ordered
and irradiated samples of MgB$_2$ are shown in Fig.1.

\begin{figure}[bp]
\includegraphics[width=0.5\textwidth,viewport=0 300 600 790]{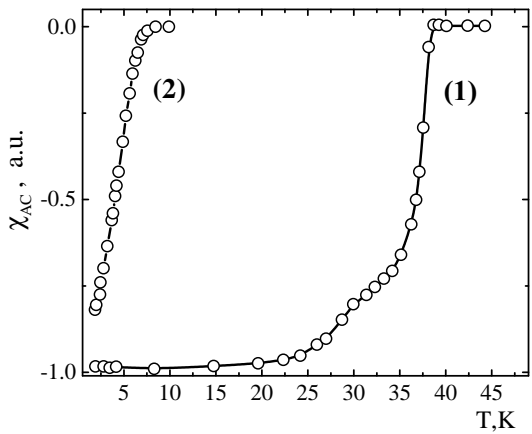}
\caption{AC-susceptibility superconducting transition curves for ordered 
        (1) and $n^1$-irradiated (2) samples of MgB$_2$.}
\label{fig1}
\end{figure}

NMR measurements were carried out on a pulse spectrometer over the 
temperature range of 20-300K in magnetic field 9.1T. The spectra of $^{11}$B and 
$^{25}$Mg were obtained by Fourier transformation of the second half of the spin-echo 
signal followed the $(\pi /2)_x- \tau _{del} -(\pi)_x$ pulse sequence. The broad 
spectra exceeding the frequency band excited by rf-pulse were measured by 
summation of an array of Fourie-signals of an echo accumulated at different 
equidistant operating frequencies. The components of magnetic shift tensor 
($K_{iso}$, $K_{ax}$) as well as the electric field gradient (EFG) parameters - 
quadrupole frequency $\nu _Q$ and asymmetry parameters $\eta$ - were determined 
by computer simulation of the measured NMR spectra. The powder pattern 
simulation program takes into account the quadrupole coupling corrections up to the 
second order of the perturbation theory. The water solution of MgCl$_2$ was used as 
a reference for $^{25}$Mg and liquid standard BF$_3$OEt$_2$ as a reference for $^{11}$B. 

Nuclear spin-lattice relaxation rates $T_1^{-1}$ of $^{11}$B and $^{25}$Mg were 
measured using saturation technique and then we deal with Fourie-transformed signal 
of echo. Saturating rf comb included itself about 50 pulses. Duration of the pulse and 
frequency were randomly changed from pulse to pulse. The range of variations in 
frequency $\delta \nu$ was taken about the total width of the NMR spectrum 
including all of transitions for powder sample ($^{25} \delta \nu \sim 1$MHz, $^{25} 
\delta \nu \sim 2$MHz).  Being applied during an interval $t<<T_1$ this broad-band 
saturating sequence of rf pulses equalizes nicely populations of magnetic levels in the 
spin systems of $^{11}$B($^{11}$I=3/2) and $^{25}$Mg($^{25}$I=5/2) showing the nonequidistant 
energy splitting in magnetic field. As a result we obtained that in ordered sample of 
MgB$_2$ (see Fig.2) the recovery of nuclear magnetization $M_Z(t)$ follows to the 
single-exponential law: $M_Z(t) \sim (1 - exp(-t/T_1))$.  Using this experimental 
procedure for $T_1$-measurements in irradiated sample we have an opportunity to 
analyze the distribution in NSLLR arisen due structural disorder. Indeed in irradiated 
MgB$_2$ (see Fig.2) a part of the $^{11}$B nucleus $(\sim(0.07-0.08)$ of total 
amount) shows the rate of spin-lattice relaxation reduced in a factor of 10 as 
compared with the ordered sample. It is believed that these slow-relaxing nucleus 
might belong to boron sited in neighborhood of the $n^1$-tracks where the short-range 
atomic order is much more influenced in comparison with atoms of B sited in the base volume fraction of the irradiated MgB$_2$. 

\begin{figure}[!]
\includegraphics[width=0.5\textwidth,viewport=0 360 600 790]{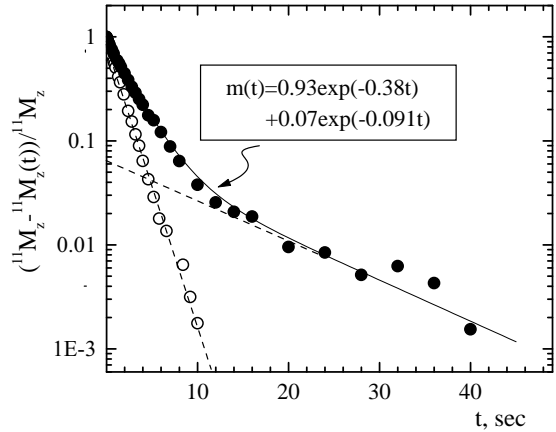}
\caption{Recovery of  $^{11}$B nuclear magnetization $^{11}M_Z(t)$ shown
in semi-log scale of the quantity $^{11}m(t)=(M_Z(\infty) - M_Z(t)) / M_Z(\infty)$
versus time interval t after applying broad-band saturating comb of rf
pulses to the ordered (open circles) and irradiated (closed circles)
samples of MgB$_2$($T=100$K).}
\label{fig2}
\end{figure}

\subsection{$^{25}$Mg NMR}

  Magnetic shift of $^{25}$Mg NMR line is found as positive isotropic quantity in 
normal state of the ordered MgB$_2$. Its value $^{25}K_{iso}$ =280(30) ppm does not 
change with temperature in the range 20-300K ($T_c$(9T)$<$15K \cite{Karkin}). In 
NSLRR measurements of $^{25}$Mg it was found that product $^{25}(T_1T)$ is also 
independent of temperature above $T_c$ and equals $^{25}(T_1T)$=350(30) sec$\cdot$K.
Magnetic shift of $^{25}$Mg in MgB$_2$ consists of two base parts, the 
chemical shift $^{25} \delta$ and the Knight shift $^{25}K_{sp}$. As shown in 
\cite{Pavarini} the Fermi-contact interaction  of $^{25}$Mg with carriers of conduction 
band should dominate both in $^{25}K_{sp}$ and  $^{25}T_1^{-1}$. The independent of 
temperature behavior of $^{25}K_{sp}$  and  $^{25}T_1^{-1}$ is in favor of this prediction 
and demonstrate that DOS near $E_F$ is flat at the energy scale exceeding 0.1eV. 
Applying for $s$-states of Mg an approximation of the free electron gas  we estimated 
$^{25}K_{sp}$ = 470 ppm using appropriate form of the Korringa relation \cite{Narath}

\begin{eqnarray}
^{25}(T_1TK_{sp}^2)S=2\mu_B^{2}/^{25}\gamma^2hk_B; S=1 
\end{eqnarray}

Correspondingly the chemical shift of $^{25}$Mg in ordered MgB$_2$ might be 
estimated as $^{25}\delta= - 190$ppm.

In irradiated sample total NMR shift decreases slightly towards 
$^{25}K_{iso,irrad}$ =240(30) ppm and within accuracy of $^{25}T_1$-measurements the 
product $^{25}(T_1T)$=360(80) sec$\cdot$K is nearly the same as before irradiation. This means 
that for the base part of the sample the contribution of  Mg(s) states to the total DOS 
is influenced negligibly by structural disorder induced during $n^1$-irradiation .

\subsection{$^{11}$B NMR}

\begin{figure}[!]
\includegraphics[width=0.5\textwidth,viewport=0 0 600 790]{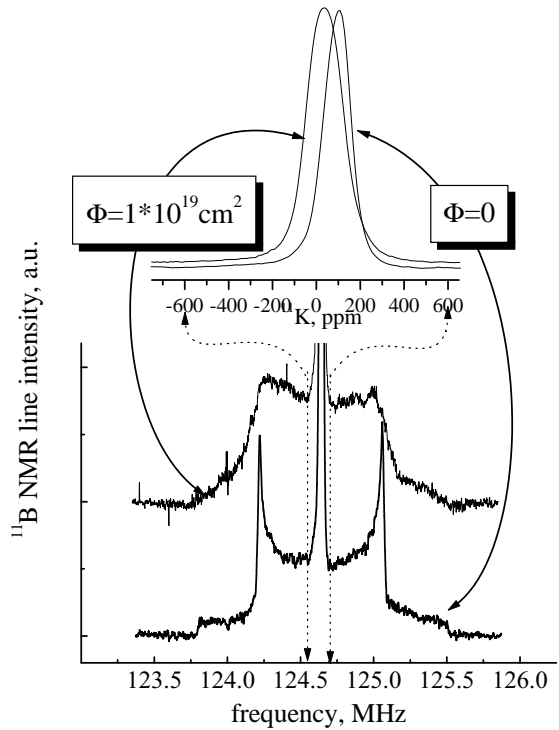}
\caption{ $^{11}$B NMR spectra of ordered ($\Phi=0)$ an $n^1$-irradiated
        ($\Phi=1 \cdot 10^{19}$cm$^{-2}$) MgB$_2$ measured in the magnetic
        field of 9.123T at T=100K.} 
\label{fig3}
\end{figure}

The NMR spectra of $^{11}$B including all transitions are shown in Fig.3 for 
$T=100$K both for ordered ($\Phi=0$) and irradiated
($\Phi=1 \cdot 10^{19}$cm$^{-2}$) samples of 
MgB$_2$. The spectrum is additionally broadened under irradiation due to the 
distribution of magnetic shift  $(\Delta K \approx 40$ ppm for central line)
and quadrupole frequency ($\Delta\nu_Q\approx$100 KHz for satellite lines).
Spherical components of the magnetic shift ($^{11}K_{iso}$, $^{11}K_{ax}$)
and the EFG ($^{11}\nu_Q$,$^{11}\eta$) tensors were determined by simulating
the powder patterns NMR line shape and data obtained are listed in Table 1. 
The measurements at different $T$ have demonstrated that magnetic 
shift and EFG parameters are independent of temperature in normal state
down to $T_c$(H) for both  structural states of MgB$_2$.
Substantial decrease of $^{11}K_{iso}$ occurs in irradiated sample,
$^{11}K(\Phi=0)-^{11}K(\Phi=1 \cdot 10^{19}$ cm$^{-2}$) =70 ppm.

The total NMR line shift of $^{11}$B might be separated in two parts. 

\begin{eqnarray}
^{11}K =^{11}\delta +^{11}K_{sp}				
\end{eqnarray}

The chemical shift $^{11}\delta$ is due to hyperfine magnetic fields originating from 
the orbital motion of electrical charges in applied external magnetic field. The Knight 
shift $^{11}K_{sp}$ is determined  by hyperfine interactions of nuclei with magnetic 
moments associated with the electron spin of carriers in the conduction band 
\cite{Slichter}.
Here we consider only an isotropic part $^{11}K_{sp,iso}$, of the 
Knight shift, which is contributed by  the Fermi-contact interaction
with conducting $s$-electrons ($^{11}K_{s}$) and the hyperfine core
polarization effects ($^{11}K_{cp}$) of B$(2p)$-electrons occupying states near $E_F$. 
\begin{eqnarray}
^{11}K_{sp,iso}=^{11}K_{s}+^{11}K_{cp}
\end{eqnarray}

Either of these contributions is proportional to the density of states in corresponding 
bands.

According the $^{25}$Mg NMR data the partial DOS of the Mg(s)-states
is roughly  unchanged by induced disorder. It is reasonable to assume
that the Fermi-contact term $^{11}K_{s}$ is also insensitive to to 
the gain of  structural disorder considered in this work

An estimate of the core-polarization Knight shift was performed 
in ~\cite{Pavarini} as a result of $ab$ $initio$ calculations and it was
shown that $^{11}K_{cp,calc}$ =-7 ppm is negative and
$|^{11}K_{cp}|<<^{11}K_{s} \approx 30$ ppm for boron in MgB$_2$. We are
far from the opinion that real magnitude of $|^{11}K_{cp}|$ should deviate 
strongly from the calculated in ~\cite{Pavarini}. 

Induced disorder might influence paramagnetic contribution
$^{11}\delta_{orb}$ to the total chemical shift of $^{11}$B. The term
$^{11}\delta_{orb}$ arises as the second-order
perturbation effect of magnetic field  on the orbital motion of
electrons in partially filled $2p_{x,y}$ bands. Crude estimate of this term
$^{11}\delta_{orb}\approx 2\mu_B<r^{-3}>_{2p}/W\sim$~200ppm might be
obtained using  $<a_0/r^3>_{2p}$ =1.1 \cite{Pavarini}  and $W\sim$~2eV
($a_0$ is the Bohr radius and $W$- the width of $2p_{x,y}$ conducting
band near $\Gamma$-A line \cite{Kortus,Medvedeva,An}.
An increase of $W$ due the induced disorder might 
result in decrease of $^{11}\delta_{orb}$ which, as believed,
determines in main the variation of $^{11}K$ under irradiation.
Of course in support of this assumption it will be desirable to
have the much more precise estimates for $^{11}\delta_{orb}$ 
obtained as a result of the band calculations.

\begin{figure}[!]
\includegraphics[width=0.5\textwidth,viewport=0 350 600 790]{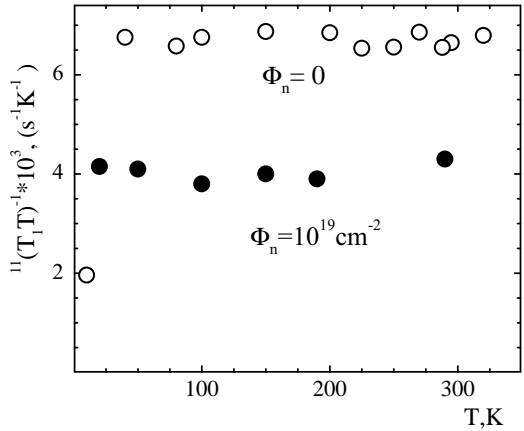}
\caption{The temperature dependence of $(^{11}T_1T)^{-1}$ product vs $T$
 in the ordered $(\Phi=0)$ and $n^1$-iradiated
 $(\Phi=1 \cdot 10^{19}$cm$^{-2}$) MgB$_2$.}
\label{fig4}
\end{figure}

It is more reliably to trace the reduction in partial DOS of
$2p$-states in irradiated MgB$_2$ considering the $^{11}T_1$
data (see Fig.4). In the normal state the product $^{11}(T_1T)$
is found as independent of temperature both in the ordered
($^{11}(T_1T)$= 155(5) sec$\cdot$K) and irradiated
($^{11}(T_1T)$ =250(20) sec$\cdot$K) samples. As for 
$^{25}$Mg  an independent of temperature product $^{11}(T_1T)^{-1}$
may be considered as an evidence for a flatness of the $N(E)$
curve near the Fermi energy at the scale of $\sim$1000K.

 As shown in \cite{Belashenko,Pavarini} the orbital relaxation  
mechanism dominates in spin-lattice relaxation rate of $^{11}$B in MgB$_2$. 
According to \cite{Obata}, \cite{Asada} we have  the following expression for 
isotropic part of $(^{11}T_1T)_{orb}^{-1}$ : 

\begin{eqnarray}
(^{11}T_1T)_{orb}^{-1}= \nonumber \\
\frac{16}{3}\mu_B^{2}(^{11}\gamma)^2hk_B<r^{-3}>_{2p}
N_{2p}(E_F)^2[f_x(f_x +2f_z)]	
\end{eqnarray}

The orbital NSLRR is proportional to $N_{2p}(E_F)^2$ and depends on the orbital 
content of $2p_i$ electrons, having near the Fermi energy appropriate DOS 
$N_i=f_i*N(E_F)$.

Using expression (4) and the $(^{11}T_1T)$-data listed in Table 1 we find
that the partial $2p$ DOS near $E_F$  in irradiated MgB$_2$ is decreased
by a factor 0.75 in comparison with its magnitude in the ordered unirradiated
sample.  According the Mc Millan formula the effect of reduced
$N_{2p}(E_F)$ results in suppression of the superconducting transition 
temperature down to $T_c\approx$~10K. A little bit less 
temperature  of superconducting transition $T_c^{onset}$~=7K was revealed
in AC susceptibility measurements.

At lasts an average magnitude of the quadrupole frequency
$^{11}\nu_Q=e^{11}QV_{zz}/2h$ is found to decrease slightly 
under disorder. The boron electric 
field gradient $V_{ZZ}$ is defined by anisotropy in the
electron occupancies $p_i$  of B $2p$ bands in the following
manner~\cite{Medvedeva_2}

\begin{eqnarray}
 V_{zz}\approx p_z-(p_x+p_y)/2					
\end{eqnarray}

The reduced value of $^{11}\nu_Q$  in irradiated sample  might be
considered in favor of the redistribution in occupancies of
B $2p$ bands resulting in the more filled $2p_{x,y}$ bands of MgB$_2$
disordered by neutron irradiation.

\section{Conclusion}
An influence of structural disorder induced by $n^1$-irradiation up to the fluence 
of thermal neutrons $(\Phi=1 \cdot 10^{19}$cm$^{-2})$ on the density of the electron states (DOS) 
near the Fermi energy was studied in MgB$_2$ by measuring NMR line shift and 
nuclear spin-lattice relaxation of $^{11}$B and $^{25}$Mg. 
According the NMR data obtained

- the partial DOS of s-states is not influenced by disorder;

- substantial decrease in DOS of $2p$-states occurs under irradiation;

- an effect of the reduced $2p$-DOS dominates in suppression of the 
superconductivity in  MgB$_2$ structurally disordered by neutron irradiation.

\begin{table}[bp]
\caption{The unit cell parameters, the $^{11}$B NMR shift
         and the quadrupole coupling parameters of  boron
         and $T_c$ in the ordered and $n^1$-irradiated
         MgB$_2$}
\begin{center}
\begin{tabular}{c|c|c}\hline\hline
   MgB$_2$                  &  $\Phi=0$   & $\Phi=1 \cdot 10^{19} cm^{-2} $ \\ \hline\hline
   a,nm                     & 0.30878   & 0.30953                         \\
   c,nm	             &0.35216	 & 0.35533                         \\
   a/c	                     &1.140	 & 1.148                           \\
   $T_c^{onset}$,K           &38	 & 7                               \\
   $^{11}K_{iso}$,ppm       & 100(10) \cite{Rem} &30(15)                  \\
   $^{11}K_{ax}$,ppm        &30(5)  	 &0-20                             \\
   $^{11}\nu_Q$,KHz         &828(10)	 &790(20)                          \\
   $^{11}\eta$	             &0.0        &0.0-0.1                         \\
   $^{11}T_1T$,sec$\cdot$K  &155(5)	 &250(20)                         \\
                                                                          \\
\end{tabular}
\end{center}
\end{table}

\begin{acknowledgments}
The authors are grateful to Dr.V.G.Zubkov,  Dr.T.V.D'yachkova and Dr.A.P.Tyutyunnik. 
for providing sample of MgB$_2$. Work is supported by the Russian State contract 
No107-1(00)-P/order No22/ and Russian State Program of Support of Leading 
Scientific Schools ( project No 00-15-96581).
\end{acknowledgments}


\begin{references}
\bibitem{Rem}
In previous paper \cite{Gerashenko} the diamond form of BN was used as a 
reference.  In measurements at room temperature the peak of  $^{11}$B NMR line in this 
solid reference is shifted down at 75ppm relative to the liquid standard 
BF$_3$OEt$_2$ used in present work and at 85ppm relative to H$_3$BO$_3$ used as 
$^{11}$B reference in \cite{Jung}.
\end{references}
\end{document}